\documentclass[10pt, conference]{IEEEtran}
\usepackage{amsfonts,subfigure,color,amsbsy}
\usepackage[nolist]{acronym}
\usepackage{graphicx,cite,amssymb,amsthm,bm}
\usepackage{amsmath}
\usepackage{mathrsfs}
\usepackage{multirow}
\usepackage{bigstrut}
\usepackage{psfrag}
\usepackage{stackengine}
\usepackage{xcolor}
\usepackage{mathtools}
\usepackage{tabularx,array}
\usepackage{mathbbol}
\usepackage{lipsum}
\usepackage[sans]{dsfont}
\usepackage{array, makecell}
\usepackage{float}
\usepackage{balance}
\usepackage{verbatim}
\usepackage{mathrsfs}
\usepackage{tikz}
\usepackage{pgfplots}
\usepackage{scalerel}
\usepackage{multirow}
\usepackage{graphicx}
\usepackage{pgfplots}
\usepackage{tikz,bm,scalerel}
\usepackage{amssymb}
\usepackage{amsfonts}
\usepackage{mathrsfs}
\usepackage{fancyref}
\usepackage{textcomp}
\usepackage{color}
\usepackage{bbm}
\usepackage{xspace}

\usepackage{tikz}

\input{style_Sauter.sty}

\usetikzlibrary{shapes,arrows}
\usetikzlibrary{calc}

\renewcommand{\mathbf}{\bm}

\newcommand{\indicator}{\mathbb{1}}
\newcommand{\dec}{\text{\rm \texttt{DEC}}}

\newcommand{\code}{\mathcal{C}}

\newcommand{\fieldtwo}{\mathbb{F}_{\!2}}

	\newtheorem{thm}{Theorem}
	
	\newtheorem{rem}{Remark}
	
	\newtheorem{dfn}[thm]{Definition}

\begin{acronym}
	\acro{AWGN}{additive white Gaussian noise}
	\acro{biAWGN}{binary-input additive white Gaussian noise}
	\acro{B-DMC}{binary-input discrete memoryless channel}
	\acro{BCJR}{Bahl-Cocke-Jelinek-Raviv }
	\acro{BP}{belief propagation}
	\acro{BPSK}{binary phase shift keying}
	\acro{BSC}{binary symmetric channel}
	\acro{BEC}{binary erasure channel}
	\acro{BP}{belief propagation}
	\acro{DE}{density evolution}
	\acro{GA}{Gaussian approximation}
	\acro{i.i.d.}{independent and identically distributed}
	\acro{DMC}{discrete memoryless channel}
	\acro{LDPC}{low-density parity-check}
	\acro{LHS}{left-hand-side}
	\acro{ML}{maximum likelihood}
	\acro{r.v.}{random variable}
	\acro{PEP}{pairwise error probability}
	\acro{p.m.f.}{probability mass function}
	\acro{p.d.f.}{probability density function}
	\acro{CCC}{constant composition code}
	\acro{SNR}{signal-to-noise ratio}
	\acro{EXIT}{extrinsic information transfer}
	\acro{PEXIT}{protograph extrinsic information transfer}
	\acro{VN}{variable node}
	\acro{CN}{check node}
	\acro{FER}{frame error rate}
	\acro{MN}{MacKay-Neal}
	\acro{MI}{mutual information}
	\acro{RA}{repeat-accumulate}
	\acro{RHS}{right-hand-side}
	\acro{RV}{random variable}
	\acro{NS}{non-systematic}
	\acro{UER}{undetected error rate}
	\acro{UEP}{undetected error probability}
	\acro{SC}{successive cancellation}
	\acro{GSCL}{generalized successive cancellation list}
	\acro{SCL}{successive cancellation list}
	\acro{TEP}{total error probability}
	\acro{ASCL}{augmented successive cancellation list}
	\acro{CRC}{cyclic redundancy check}
	\acro{RCB}{random coding bound}
	\acro{eMBB}{enhanced mobile broadband}
	\acro{mMTC}{massive machine-type communication}
	\acro{URLLC}{ultra-reliable low-latency communications}
	\acro{RCU}{random coding union}
\end{acronym}

\IEEEoverridecommandlockouts
\begin{document}
\title{Error Detection Based on Generalized Successive Cancellation List Decoding for Polar Codes}
\author{\IEEEauthorblockN{Alexander Sauter\IEEEauthorrefmark{1}\IEEEauthorrefmark{3},
		Mustafa Cemil Co\c{s}kun\IEEEauthorrefmark{2}, and
		Gianluigi Liva\IEEEauthorrefmark{1}}
	\IEEEauthorblockA{\IEEEauthorrefmark{1}\textit{German Aerospace Center,
			We{\ss}ling, Germany}}
	\IEEEauthorblockA{\IEEEauthorrefmark{2}\textit{Nokia Bell Labs, Murray Hill, New Jersey, US }}
	\IEEEauthorblockA{\IEEEauthorrefmark{3}\textit{Institute for Communications Engineering, Technical University of Munich, Munich, Germany}} 
	\{alexander.sauter, gianluigi.liva\}@dlr.de, mustafa.coskun@nokia-bell-labs.com
	% <-this % stops an unwanted space
	\thanks{Alexander Sauter and Gianluigi Liva acknowledge the financial support by the Federal Ministry for Research, Technology and Space (BMFTR) of Germany in the programme of ``Souver\"an. Digital. Vernetzt." Joint project 6G-RIC, project identification number: 16KISK022.
}}

\maketitle
\IEEEoverridecommandlockouts
\begin{abstract}
	Successive cancellation list (SCL) decoding has been widely adopted for polar codes, which allows near \acl{ML} performance with sufficiently large list size. In this work, we show that, if the list size is $2^{\gamma}$, where $\gamma$ is the fundamental quantity called mixing factor, then a modification to SCL decoding can implement Forney's generalized decoding rule. Hence, it provides an efficient means to discard unreliable decisions. The performance achieved by short polar codes under the proposed generalized SCL decoding is analyzed via Monte Carlo simulations.
\end{abstract}

\begin{IEEEkeywords}
	Ultra-reliable low-latency communication, polar codes, SCL decoding, generalized decoding, error detection
\end{IEEEkeywords}

\section{Introduction}\label{sec:intro}

Polar codes \cite{Ari09} made their way in the fifth generation of mobile communication standard (5G NR) as the error correction scheme for the control channel  \cite{3GPP21}. This result is mainly due to their good performance in the short blocklength regime, when used in concatenation with an outer high-rate binary block code, and with low-complexity \ac{SCL} decoding \cite{TV15}. 

Among its numerous applications, the 5G NR standard targets \ac{eMBB} and \ac{URLLC}. \ac{URLLC} cover, in particular, mission-critical applications with a demand for a strong protection of data packets. On the physical layer of the communication systems, reliability is measured by the frequency of decoding errors. Decoding errors can lead to undesired behavior, particularly for mission-critical applications sensitive to \emph{undetected errors} (see, e.g., \cite{Dolinar2008:Bounds}). An undetected error occurs when the decoder outputs a valid codeword different from the transmitted one; hence, the upper layers of the protocol stack are not informed of the incorrect decision. In mission-critical applications, it is, hence, imperative to design codes and decoding algorithms that can enable precise control of the \ac{TEP} and the \ac{UEP}.
A pragmatic approach to perform error detection is to concatenate an (inner) error correcting code with an (outer) error detection code, where the latter is typically a \ac{CRC} code. The role of the error detection code is to validate the decision at the output of the inner code decoder. This approach, which works well for large data packets, becomes increasingly inefficient as the blocklength reduces \cite{Sauter25}, due to the non-negligible rate loss caused by the inclusion of the outer error detection code.
In \cite{For68}, Forney introduced a generalized decoding rule that allows one to achieve the optimum tradeoff between \ac{TEP} and \ac{UEP}. The rule relies on a threshold test, that can be applied to the output of a \ac{ML} decoder \cite{Hof2009:ISTCA}. An important consequence of the approach outlined in \cite{For68} is that optimum error detection can be achieved without the need to introduce an outer error detection code. However, the generalized decoding rule of \cite{For68} requires the evaluation of a metric that  can be performed efficiently for only certain classes of linear block codes, e.g., terminated convolutional codes \cite{Raghavan1998,Hof2009:ISTCA,Williamson2014:ROVA,Wesel2022:ROVA}, whereas, for a general $(n, k)$ binary linear block code, the complexity grows with the minimum between $2^k$ and  $2^{n-k}$.

In \cite{Fazeli2021, YFV23:wef_polar}, the notion of mixing factor $\gamma$ (that is, the number of information bits preceding the last frozen bit, at the polar transform input) is defined as one of the most fundamental parameters for polar codes. In \cite{YFV23:wef_polar}, an algorithm is introduced to compute the exact distance spectrum of any polar code, whose complexity scales exponentially with the mixing factor. At the same time, \cite{Fazeli2021} shows that \ac{ML} decoding is achieved by a modified \ac{SCL} decoding with list size $2^{\gamma}$. In this paper, we show that Forney's decoding metric can be evaluated exactly by an \ac{SCL} decoder with list size $2^{\gamma}$; hence, the optimum generalized decoding is achieved via such modified SCL decoding. When $\gamma \leq n-k$, this decoding strategy can significantly reduce complexity compared to a brute force evaluation of Forney's metric.\footnote{Note that $\gamma$ is always upper bounded by $k$.}Leveraging this result, we propose a \ac{GSCL} decoding algorithm that implements the threshold test of \cite{For68}, which mimics the optimum generalized decoding rule. Our results complement the analysis of \cite{Yuan24}, where a clever modification of the \ac{SCL} decoder is proposed that allows to tightly approximate the optimum metric of \cite{For68}.

The rest of the paper is organized as follows. Section~\ref{sec:preliminaries} introduces the notation, the fundamentals of polar codes, and Forney's generalized decoding rule. Section~\ref{sec:decRule} introduces \ac{GSCL} decoding of polar codes. Numerical results are provided in Section~\ref{sec:results}. Section~\ref{sec:conclusions} concludes this work.

\newcommand{\modulate}[1]{{\mathsf{B}\left(#1\right)}}

\section{Preliminaries}\label{sec:preliminaries}
\subsection{Notation}\label{suc:notation}
We denote length $n$ vectors as $x^n = [x_1, \dots, x_n]$, and sets by capital calligraphic letters, e.g., $\mathcal X$. The cardinality of a set is denoted by $\abs{\mathcal X}$. The $i$th vector of a set is denoted by small bold letters with a lowercase number, e.g., $\bm x_i$. We use the notation $x_a^{b}$ as shorthand to denote a vector $[x_a, \dots, x_b]$, where the $i$th entry of the vector is denoted by $x_i$. Matrices are denoted by capital bold letters, e.g., $\bm G$. Uppercase letters refer to \acp{RV}, and small counterparts to their realizations. We write $\log(\cdot)$ to denote the natural logarithm, and $\log_2(\cdot)$ to denote the base-$2$ logarithm. We use $\fieldtwo$ to denote the binary finite field. Moreover, $\indicator(\cdot)$ is the indicator function, and $[n]$ is used for the set $\{1,2,\ldots,n\}$. A \ac{B-DMC} is defined as $W: \mathcal{X} \rightarrow \mathcal{Y}$ with input alphabet $\mathcal{X} = \{0,1\}$, and arbitrary output alphabet $\mathcal{Y}$. The channel transition probability for $n$ independent channel uses is denoted by $W^n(y^n| x^n) = \prod^n_{i=1} W(y_i|x_i)$.

\subsection{Polar Codes}\label{sec:polar_enc}
Let the binary polar transform be the $n\times n$ matrix 
%
%\Mod{
	\begin{align}\label{eq:polar_trafo}
		\bm G_n= \bm{G}_2^{\otimes m}
	\end{align}
	%
	% where $n=\log_2 N$  and
	% \begin{align}
		% \bm F=\begin{bmatrix}
			% 1 & 0 \\
			% 1 & 1
			% \end{bmatrix}
		% \end{align}
	%
	where $n=2^m$ with a non-negative integer $m$ and the superscript ${\otimes m}$ denotes the $m$-fold Kronecker product of the polarization kernel 
	\begin{equation}
		\bm G_2 = \begin{bmatrix}
			1 & 0\\
			1 & 1
		\end{bmatrix}.
	\end{equation}
	%and $m=\log n$. %\bm B_N$ is the bit-reversal matrix described in \cite{Ari09}. 
	An $(n,k)$ polar code of length $n$ and dimension $k$ is defined by set $\mathcal{A} \subseteq [n]$ with cardinality $|\mathcal{A}|$ being the number of information bits. Equivalently, it can be defined by the complementary set $\csetA = [n] \setminus \mathcal{A}$, which is the set containing the indices of \emph{frozen} bits. The code rate is $R=k/n$.
	Encoding is performed as $x^n=u^n\bm{G}_n$, where $u_i$ carries an information bit if $i\in\mathcal{A}$, or it is set to a predefined value if $i \in \csetA$ (frozen bit).
	For a given channel, the set $\mathcal{A}$ can be determined, for example, via \ac{DE} analysis \cite{richardson_capacity} by selecting the synthetic channel indices in $[n]$ with the highest reliability under genie-aided \ac{SC} decoding \cite{Ari09,Mori2009,Tal2013}. Throughout the paper, we denote by $s = \max (\csetA)$ the index of the last frozen bit.
	
	\subsection{Successive Cancellation List Decoding}
	\ac{SC} decoding decides the values of the bits $u_1, u_2, \ldots, u_n$ sequentially. In particular, for the $i$th bit, \ac{SC} decoding computes the term $W_n^{(i)}(y^n, \hat{u}^{i-1}|u_i)$ based on the channel output $y^n $ and the previously decoded bits $\hat{u}^{i-1}$. Then, a decision on $u_i$ is taken by computing the logarithmic ratio
	\begin{equation} \label{eq:softdec}
		\ell_i\left(y^n,\hat{u}^{i-1}\right) = \log \frac{W_n^{(i)}( y^n, \hat{u}^{i-1}|u_i=0)}{W_n^{(i)}(y^n, \hat{u}^{i-1}|u_i=1)} 
	\end{equation}
	with the hard decision
	\begin{equation} \label{eq:harddec}
		\hat{u}_i =\begin{cases}
			u_i, & \text{if $u_i$ is a frozen bit}  \\
			0, & \text{if $\ell_i\left(y^n,\hat{u}^{i-1}\right) \geq 0$}\\
			1, & \text{otherwise}. 
		\end{cases} 
	\end{equation}
	%Erroneous decision i.e. $\hat{u}_i \neq u_i$, for $i \in \mathcal{A}$ can not be corrected in later decoding steps.
	Unlike \ac{SC} decoding, \ac{SCL} decoding does not make a hard decision on $u_i$, $i\in \mathcal{A}$, after every decoding step but pursues the decoding by keeping two decoding \emph{paths} for $\hat{u}_i = 0$ and $\hat{u}_i = 1$. To limit the complexity of \ac{SCL} decoding, one may restrict the number of threads (list size) to $L$. If the number of decoding paths exceeds $L$, the list will be pruned by keeping the $L$ most-likely paths. 
	At the $n$th decoding step, \ac{SCL} decoding selects from the final list the path with largest likelihood, outputting the corresponding decision $\hat{u}^n$ (or, equivalently, its associated codeword $\hat{x}^n$).
	
	\subsection{Mixing Factor}\label{sec:mixing}
	Fazeli, Yao and Vardy introduced in \cite{Fazeli2021 ,YFV23:wef_polar} the notion of \emph{mixing factor} of a polar code. This fundamental parameter determines the required list size for a suitably-modified version of \ac{SCL} decoding to perform \ac{ML} decoding \cite{Fazeli2021}. In addition, any polar code can be represented as a union of disjoint subsets, where the number of subsets in this representation is determined by the same parameter, i.e., the mixing factor. Note that such a representation is very useful for computing the exact distance spectrum of any polar code\cite{YFV23:wef_polar}. Given an $(n, k)$ polar code, the mixing factor is defined as
	\begin{equation}
		\gamma = k - (n - \max(\csetA)) \label{eq:mixingfactor}
	\end{equation}
	i.e., it is the number of information bits that precede the last frozen bit in the vector $u^n$. The list size for achieving \ac{ML} decoding given by 
	\begin{equation}
		L = 2^\gamma. \label{eq:listsize}
	\end{equation}
	Surprisingly, an entire polar code can be represented as the union of $2^\gamma$ disjoint subsets. Notably, setting the list size according to \eqref{eq:listsize} allows \ac{ML} decoding with standard \ac{SCL} decoding over the \ac{BEC} \cite{Hashemi17}. Over general symmetric \acp{B-DMC}, \ac{SCL} decoding with list size as in \eqref{eq:listsize} allows to closely approach the performance of an \ac{ML} decoder  \cite{Fazeli2021}. Furthermore, \ac{ML} decoding can be attained by modifying the \ac{SCL} decoder so that, for the list of paths obtained after the last frozen bit, the corresponding  most likely extension is found by means of a nearest coset decoding algorithm \cite{Fazeli2021}.

	\subsection{Forney's Generalized Decoding Rule}\label{sec:forney}
	
	Forney~\cite{For68} introduced a decoding rule, where the decoder outputs all codewords $x^n \in \mathcal{C}$ that pass the threshold test
	\begin{equation}
		\frac{W^n(y^n|x^n)}{\sum_{\tilde{x}^n \in \code\setminus x^n} W^n(y^n|\tilde{x}^n) } \geq 2^{nT} \label{eq:forneytest}
	\end{equation}
	where $T$ is a decoder parameter. If no codeword satisfies \eqref{eq:forneytest}, the decoder declares a detected error (\emph{erasure}). For $T>0$, the generalized decoding rule defined by the threshold test divides the channel output space into disjoint decision regions, resulting in either a unique codeword fulfilling \eqref{eq:forneytest}, or in an erasure. In contrast, for $T\leq 0$ the decoding regions may overlap. Hence, depending on the channel observation $y^n$, Forney's generalized decoder may output a list of codewords.
	
	\begin{rem}[On Generalized Decoding] \label{rem:MGD}Following \cite{Hof2009:ISTCA}, we consider a modification of Forney's decoding rule, which ensures unique, incomplete decoding for any value of $T$. The modification works as follows. A complete decoder  $\dec: \mathcal{Y}^n \mapsto \mathcal{C}$ computes a decision $\hat{x}^n=\dec(y^n)$. The decision is tested via \eqref{eq:forneytest}. If the decision passes the test, then decoding outputs $\hat{x}^n$. Otherwise, an erasure is declared. As observed in \cite{Hof2009:ISTCA}, if $\dec$ is \ac{ML}, the decoder coincides with Forney's generalized decoding rule for positive values of $T$. For $T\leq 0$, this decoding rule differs from Forney's one since, at most, only one codeword can be provided at its output. For $T = -\infty$, the decoding rule results in a complete decoder (with no erasures), which is \ac{ML} if and only if $\dec$ is \ac{ML}.
	\end{rem}
	
	By testing a complete decoder output, this decoding rule can be used to perform post-decoding error detection, which guarantees the optimum trade-off between \ac{UEP} and \ac{TEP} under ML decoding \cite{For68}. The threshold test may be rewritten as 
	\begin{equation}
		\frac{W^n(y^n | \hat{x}^n)}{P_{Y^n}(y^n)} \geq 2^k \frac{2^{nT}}{1 + 2^{nT}}  \label{eq:ForneyBay}
	\end{equation}
	where 
	\begin{equation}
		P_{Y^n}(y^n) = \sum_{x^n \in \fieldtwo^n} W_n(y^n | x^n) P_{X^n}(x^n; \code) \label{eq:PY}
	\end{equation}
	is the output distribution induced by the $(n,k)$ code $\code$
	\begin{align}
		P_{X^n}(x^n; \code) = \frac{1}{2^k} \mathbb{1}(x^n \in \code). \label{eq:PX}
	\end{align}
	In the following, we refer to \eqref{eq:PX} and to \eqref{eq:PY} as \emph{codebook-induced} input and output distributions, respectively.
	The evaluation of \eqref{eq:PY} is generally difficult, as it requires summing the likelihoods of all codewords. As a notable exception, \eqref{eq:PY} can be efficiently evaluated for terminated convolutional codes by computing the forward recursion in a \ac{BCJR} decoder \cite{Hof2009:ISTCA}.
	
\section{Generalized Successive Cancellation Decoding}\label{sec:decRule}
In this section, we propose a \ac{GSCL} decoding rule for polar codes that relies on the test \eqref{eq:ForneyBay}. The decoder exploits the intermediate calculations of an \ac{SCL} decoder to efficiently compute the denominator in \eqref{eq:ForneyBay}, upon fixing the list size according to \eqref{eq:listsize}.

Let us denote the uniform distribution over $\fieldtwo^n$, that assigns equal probability to all possible sequences, as
\begin{align}
	Q(x^n) = \frac{1}{2^n} \qquad \forall x^n \in \fieldtwo^n. \label{eq:QX}
\end{align}
From the codebook-induced input distribution, we have
\begin{align}
	P_{ X^n}(x^n; \code) = 2^{(n-k)} Q(x^n) \indicator(x^n \in \code), \label{eq:incudedInput}
\end{align}
Substituting \eqref{eq:incudedInput} into \eqref{eq:PY}, we obtain 
\begin{align}
	P_{Y^n}(y^n) = 2^{(n-k)}\sum_{x^n \in \code} W^n(y^n | x^n) Q( x^n). 
\end{align}
Now, suppose the codebook $\code$ is partitioned into $L$ disjoint subsets $\code_1, \code_2, \ldots, \code_L$, with 
\begin{equation}
	\code = \bigcup^L_{i=1} \code_i 
\end{equation}
and  $\code_i \cap \code_j =  \emptyset$ for $i \neq j$.
Let us denote by
\begin{equation}
	W^n(y^n,x^n) = W^n(y^n|x^n)Q(x^n)
\end{equation}
the joint distribution of $Y^n$ and $X^n$, induced by the uniform distribution \eqref{eq:QX}.
The codebook-induced output distribution  can be expressed by the sum of the $L$ contributions
\begin{equation}
	P_{Y^n}(y^n)   = 2^{(n-k)} \sum_{l= 1}^L \sum_{x^n \in \code_l} W^n(y^n, x^n).  \label{eq:OutputPUnion}
\end{equation}

Consider now an $(n,k)$ polar code under \ac{SCL} decoding with list size set to $L=2^\gamma$. After processing the last frozen bit $u_s$, the $L$ paths stored in the list are defined by the $L$ vectors $$u^s\{1\}, u^s\{2\}, \ldots, u^s\{L\}.$$ Here, $u^s\{l\}$ represents the decisions taken, for the $l$th path, by the \ac{SCL} decoder up to the last frozen bit. We make use of the following definition.
\begin{dfn}[Polar codebook partition]\label{def:partition}
	We define the partition of the polar codebook as
	\begin{equation}\label{eq:partition}
		\code_l = \left\{ x^n = u^n \bm G_2^{\otimes m} : u^s = u^s\{l\}, \, u_{s+1}^n \in \fieldtwo^{n-s} \right\} 
	\end{equation}
	for $l = 1, \dots, L$.
\end{dfn}
Definition \ref{def:partition} introduces a partitioning of the polar codebook in $L$ disjoint subsets, which appears as \cite[Proposition 1]{YFV23:wef_polar}. Thus, to compute \eqref{eq:PY}, it is sufficient to use \eqref{eq:OutputPUnion} with $\code_l$ defined according to \eqref{eq:partition}. Note that, for $x^n = u^n \bm{G}_2^{\otimes m}$, and owing to \cite[Equations 4 and 5]{Ari09},
\begin{equation}
	\sum_{x^n \in \code_l} W^n(y^n, x^n) = \frac{1}{2}  W_n^{(s)}\left(y^n, u^{s-1}\{l\}\, \big| \, u_s\right) \label{eq:coset}
\end{equation}
where $W_n^{(s)}(y^n, u^{s-1}\{l\}\, | \, u_s)$ is the probability-domain path metric computed recursively by the \ac{SC} decoder for the path defined by $ u^{s}\{l\}$ \cite[Proposition 3]{Ari09}. Observe that one does not have to compute the probabilities for all $|\code_l|$ members of the subset in order to compute left-hand side of \eqref{eq:coset}. Hence, \eqref{eq:PY} can be obtained as a byproduct of the \ac{SCL} decoder as

\begin{equation}
	P_{Y^n}(y^n) = 2^{n-k-1} \sum_{l=1}^L W_n^{(s)} \left(y^n, u^{s-1}\{l\} \, \big| \, u_s \right)  \label{eq:outputDist}
\end{equation}
i.e., by summing the probability-domain path metrics computed after the last frozen bit, and dividing the result by $2^{n-k-1}$.

\ac{GSCL} decoding of an $(n,k)$ polar code can be summarized as follows:

\medskip
\begin{itemize}
	\setlength\itemsep{0.5em}
	\item[A.] Evaluate $P_{Y^n}(y^n)$ via \eqref{eq:outputDist} with $L=2^\gamma$.
	\item[B.] Given a threshold $T$, the \ac{SCL} decoder decision $\hat{x}^n$, and the result of step A, perform the threshold test \eqref{eq:ForneyBay}.
\end{itemize}

\medskip
Some observations follow. First, even by setting $L=2^\gamma$, \ac{SCL} decoding does not guarantee to obtain the \ac{ML} decision. Thus, strictly speaking, the combination of \ac{SCL} decoding with the test  \eqref{eq:ForneyBay} does not implement Forney's generalized decoding rule. However, numerical results show that, by choosing $L=2^\gamma$, \ac{SCL} decoding approaches tightly the \ac{ML} decoder performance. We hence expect \ac{GSCL} to yield a near-optimum tradeoff between \ac{TEP} and \ac{UEP}, for a given polar code. Second, by employing the modified \ac{SCL} decoder of \cite{Fazeli2021}, it is possible to implement the actual generalized decoding rule of Forney. Finally, it should be noted that the complexity of \ac{GSCL} decoding grows exponentially in the mixing factor, which is strictly smaller than the code dimension. While this result may hinder the use of \ac{GSCL} decoding in many cases of interest, low-complexity accurate approximations of the right hand side of \eqref{eq:ForneyBay}  based on an ingenious tree search algorithm were recently proposed \cite{Yuan24}.

\section{Numerical Results}\label{sec:results}

\begin{figure}[t]
	\centering
	\includegraphics[width = 0.95\columnwidth]{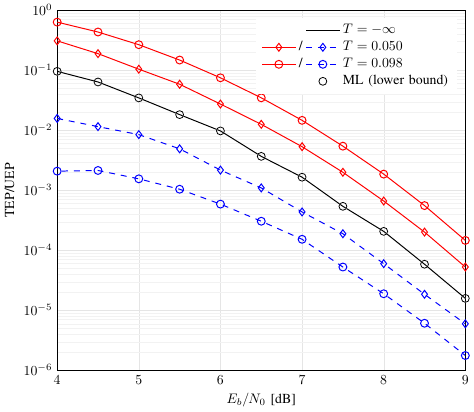}
	\caption{\ac{TEP} and \ac{UEP} versus $E_b/N_0$ for $(64,48)$ polar code with list size $L=32$. The \ac{UEP} is denoted by dashed lines, whereas the \ac{TEP} is denoted by solid lines.}
	\label{fig:VarioutT}
\end{figure}

In this section, we analyze the performance of \ac{GSCL} decoding via Monte Carlo simulations.
We consider the transmission over a \ac{biAWGN} channel. The noise variance is $\sigma^2$, and the channel \ac{SNR} is given by ${E_b}/{N_0}={1}/(2R \sigma^2)$ where $E_b$ is the energy per information bit and $N_0$ is the single-sided noise power spectral density. 
To allow low-complexity decoding, the polar code design has been constrained to limit the mixing factor. In particular, we made use of the following approach: for a given \emph{maximum} mixing factor $\gamma^\star$, we enforced the choice of the frozen bits to take place within the first $n-k+\gamma^\star$ coordinates. In particular, for a target \ac{SNR}, the frozen bit set includes the $n-k$ synthesized channels with the lowest reliability within the first $n-k+\gamma$ coordinates, with the ranking of the synthesized channels performed via \ac{DE} \cite{Ari09,Mori2009,Tal2013}. The proposed polar code construction guarantees $\gamma\leq \gamma^\star$, i.e., the actual mixing factor does not exceed $\gamma^\star$ at the expense of a possible decrease in the minimum distance. In the results reported in this section, we set $\gamma^\star=8$.
\begin{figure}[t]
	\centering
	\includegraphics[width = 0.95\columnwidth]{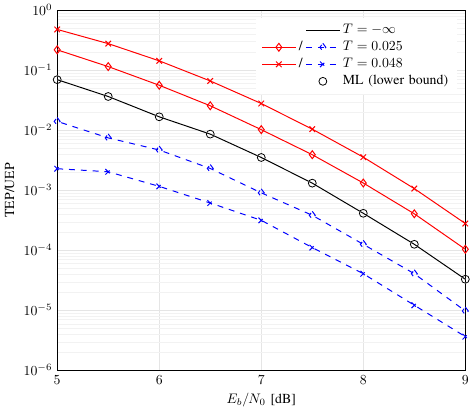}
	\caption{\ac{TEP} and \ac{UEP} versus $E_b/N_0$ for $(128,96)$ polar code with list size $L=128$. The \ac{UEP} is denoted by dashed lines, whereas the \ac{TEP} is denoted by solid lines.}
	\label{fig:ThresholdandBound}
\end{figure}
The results for a $(128,96)$ polar code are depicted in Fig.~\ref{fig:ThresholdandBound}. Here, $\gamma=7$, resulting in a list size $L=128$. The chart reports the performance for various values of the threshold $T$, including the case where $T = -\infty$, i.e., the case where no error detection is performed (all errors are, here, undetected). Note that the performance with $T = -\infty$ has been numerically found to be tight on the \ac{ML} decoder performance (the lower bound \cite{TV15} on the \ac{ML} decoder error probability is provided in the chart, and it matches the performance of the \ac{SCL} decoder). The trade-off between \ac{TEP} and \ac{UEP} is visible when the value of $T$ is chosen to be larger than zero. Fig.~\ref{fig:VarioutT} reports the results for a $(64,48)$ polar code. Here, $\gamma=5$, resulting in a list size $L=32$. As for the $(128,96)$ case, the performance with $T=-\infty$ is reported as a reference.

\section{Conclusions}\label{sec:conclusions}
In this paper, we presented a generalized successive cancellation list (GSCL) decoding for polar codes, which leverages intermediate calculations performed by a standard successive cancellation decoder to formulate the threshold test. The threshold test can be used to optimize the trade-off between total and undetected error probabilities. The approach relies on list decoding, where the list size is set according to the mixing factor of the underlying polar code.  Numerical results are presented for a polar code construction constrained to limit the mixing factor. We finally note that, although this approach can be substantially less complex compared to a brute force evaluation of Forney's generalized decoding metric, the underlying exponential dependency on the mixing factor may still result in an excessive computational cost for general polar codes, calling for less complex, sub-optimal approximations of the GSCL decoding.

%%%%%%%%%%%%%%%%%%%%%%%%%%
\balance


\end{document}